\documentclass[reprint,prl,twocolumn,longbibliography]{revtex4-1}

\usepackage{amsmath}
\usepackage{mathtools}
\usepackage{braket}
\usepackage{bbm}
\usepackage{graphicx}
\usepackage{amssymb}
\usepackage[dvipsnames]{xcolor}
\usepackage{multirow}
\usepackage{xfrac}
\usepackage{textcomp}

\usepackage{hyperref}
\usepackage{bbold}

\hypersetup{
	colorlinks=true,
	linkcolor=blue,
	citecolor=blue,
	filecolor=magenta,
	urlcolor=cyan,
	breaklinks=true
}
\DeclareMathOperator{\Tr}{Tr}

\DeclareMathOperator{\sgn}{sgn}

\DeclareMathOperator{\sech}{sech}

\newcommand{\papertitle}{Elastic backscattering of quantum spin Hall edge modes from Coulomb interactions with non-magnetic impurities}
\newcommand{\tcm}{T.C.M. Group, Cavendish Laboratory, University of Cambridge, JJ Thomson Avenue, Cambridge, CB3 0HE, U.K.}

\newcommand{\iu}{{\rm i}}
\newcommand{\dif}{\mathrm{d}}

\begin{document}
	
	\title{\papertitle}
	\author{Max McGinley}
	\affiliation{\tcm}
	\author{Nigel R. Cooper}
	\affiliation{\tcm}
	
	\date{\today}

	\begin{abstract}
		We demonstrate that electrostatic interactions between helical electrons at the edge of a quantum spin Hall insulator and a dynamical impurity can induce quasi-elastic backscattering. Modelling the impurity as a two-level system, we show that transitions between counterpropagating Kramers-degenerate electronic states can occur without breaking time-reversal symmetry, provided that the impurity also undergoes a transition. The associated electrical resistance has a weak temperature dependence down to a non-universal temperature scale. Our results extend the range of known backscattering mechanisms in helical edge modes to include scenarios where electron tunnelling out of the system is absent.
	\end{abstract}
	\maketitle

\emph{Introduction.---} The quantum spin Hall (QSH) effect \cite{Kane2005,Bernevig2006,Wu2006,Xu2006} is a prototypical example of a symmetry protected topological phase \cite{Chen2010}, featuring helical edge modes that remain gapless and conducting as long as time-reversal symmetry is maintained and the bulk gap stays open. While these edge modes have been directly observed in a variety of solid state systems using photoemission spectroscopy \cite{Konig2007}, their characteristic conductance properties are found to be much less robust than those of chiral edge modes in the integer quantum Hall effect \cite{Roth2009,Konig2013,Fei2017,Wu2018}.

Aside from time-reversal symmetry breaking due to stray magnetic fields or spontaneous magnetism \cite{Wu2018,Novelli2019,ZhaiNote}, a number of mechanisms have been put forward to account for the edge mode resistance seen in experiment. In systems with inhomogeneous doping, the bulk gap may vary with position and even close in certain regions, leading to the formation of metallic charge puddles. If helical electrons can tunnel into these gapless regions, then their protection against backscattering is compromised and the edge becomes resistive \cite{Vayrynen2013,Vayrynen2014}. This effect is particularly strong for Kramers-degenerate impurities, wherein the resultant magnetic exchange interactions can facilitate \textit{quasi-elastic} backscattering, i.e.~involving only a small energy exchange (of the order of the Kondo temperature) \cite{Maciejko2009,Tanaka2011,Altshuler2013,Cheianov2013}. 

In contrast, in the absence of such electronic exchange processes and TRS-breaking perturbations, the only known sources of resistance involve inelastic backscattering. Electron-electron and electron-phonon mediated backscattering lead to a resistance that is strongly suppressed as the temperature is decreased \cite{Schmidt2012,Budich2012,Lezmy2012,Kainaris2014}, which is inconsistent with the weak temperature dependence seen in experiment \cite{Fei2017,Wu2018}. One possible explanation is that the necessary energy is provided by external noise, giving a weaker temperature dependence \cite{Vayrynen2018}.

\begin{figure}
	\includegraphics[scale=1]{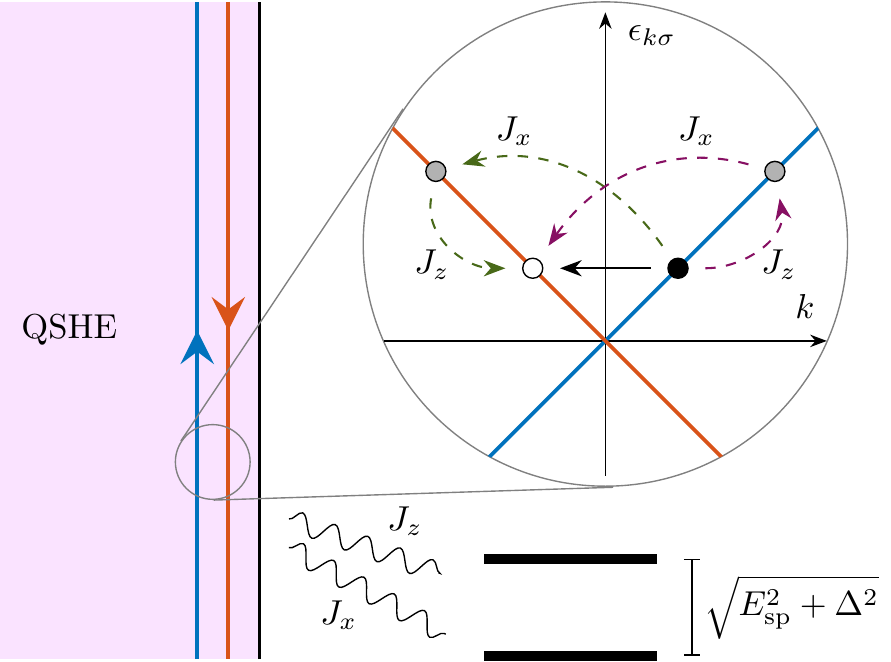}
	\caption{Quantum spin Hall insulator coupled to an impurity (modelled as a two-level system) via electrostatic interactions [Eq.~\eqref{eqCoupleBare}]. Inset: The bare couplings $J_z$ and $J_x$ [Eq.~\eqref{eqHamTerms}] combine to induce elastic backscattering between Kramers-degenerate states (black and white dots). The transition can proceed by two paths (green and violet) depending on which perturbation is applied first. Because the impurity pseudospin operators  do not commute $[\hat{\sigma}^z, \hat{\sigma}^x] = 2\iu \hat{\sigma}^y$, destructive interference of these two paths is avoided. During this process, the impurity undergoes a simultaneous transition, thus requiring an energy transfer of $\epsilon \coloneqq \sqrt{E_{\rm sp}^2 + \Delta^2}$, which can be arbitrarily small.}
	\label{figElectro}
\end{figure}

\begin{figure}
	\includegraphics[scale=1]{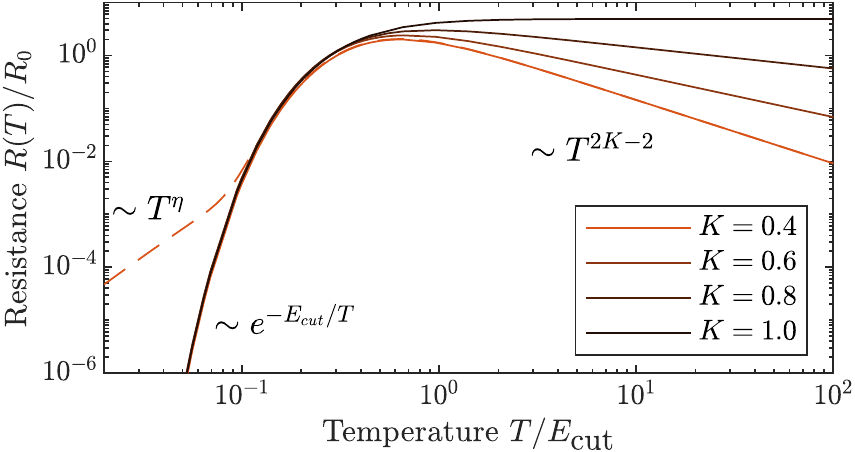}
	\caption{Resistance of helical edge modes due to interaction with a two-level system impurity, in the regime $\epsilon \gtrsim E_{y}$ such that Eq.~\eqref{eqGDC} applies with $E_{\rm cut} = \epsilon$, for various values of $K$. The resistance is plotted in units of $R_0 \coloneqq (h/e^2) \times y_0^2 (E_{\rm cut}/E_{\rm g})^{2K-2}$. For $E_{y} \gtrsim \epsilon$, the same qualitative form is expected. The dashed line includes the contributions from inelastic scattering, which are dominant for $T \ll E_{\rm cut}$.}
	\label{figRes}
\end{figure}

In this Letter, we identify a new source of edge mode resistance that does not involve tunnelling into in-gap states. Electrostatic interactions between helical electrons and a dynamic impurity (Fig.~\ref{figElectro}) are shown to induce quasi-elastic backscattering, giving a resistance profile which strongly resembles that induced by a magnetic impurity (see Fig.~\ref{figRes}): For temperatures $T$ above some non-universal cutoff $E_{\rm cut}$, the resistance scales as $T^{2K-2}$, where $K \leq 1$ is the Luttinger parameter (equal to unity for non-interacting electrons). At temperatures below $E_{\rm cut}$, the dynamics of the impurity becomes frozen, leaving only the aforementioned inelastic backscattering processes which scale as $T^\eta$ with $\eta > 0$.

Quasi-elastic backscattering is possible here because time-reversal symmetry does not act `locally' on the helical edge electrons, but on the composite system-plus-impurity, as we highlighted in Ref.~\cite{McGinley2020}.  Accordingly, transitions between Kramers-degenerate electron states -- which would be forbidden by time-reversal symmetry in the absence of any extraneous degrees of freedom -- are in fact allowed, provided that the impurity undergoes a simultaneous transition. In essence, the energy scale below which the helical edge modes are protected is not set by the electronic gap, but by the gap of the composite system, which can be arbitrarily small. Our results indicate that the elimination of charge puddles (e.g.~by using QSH insulators with larger band gaps) will not necessarily restore conductance quantization.

\emph{Setup.---}
The boundary of a two-dimensional QSH insulator hosts a single pair of counterpropagating modes in which the direction of motion is determined by the electron spin. 
In a clean, dispersionless, non-interacting system in the absence of spin-orbit coupling, the Hamiltonian can be written as $\hat{H}_0 = \sum_\sigma \sigma v_F \int \dif x\, \hat{\psi}_\sigma^\dagger(x) \iu \partial_x^{\vphantom{\dagger}} \hat{\psi}_\sigma^{\vphantom{\dagger}}(x)$, where $\hat{\psi}_{\sigma}^\dagger(x)$ creates a fermion with spin $\sigma \in \{+1,-1\}$ at a coordinate $x$ along the edge. The TRS operator $\hat{T}$ (which is antiunitary $\hat{T}\, \iu\, \hat{T}^{-1} = -\iu$) exchanges the two spin species $\hat{T} \hat{\psi}_{\sigma}(x) \hat{T}^{-1} = \sigma \hat{\psi}_{-\sigma}(x)$, and squares to $\hat{T}^2 = (-1)^{\hat{N}_F}$, where $\hat{N}_F$ is the fermion number operator. Perturbations cannot couple counterpropagating states of the same energy unless TRS is broken explicitly or spontaneously \cite{Xu2006}. This protection against elastic backscattering prevents the helical edge modes from being gapped or localized \cite{Wu2006}.

If the simple Hamiltonian $\hat{H}_0$ is supplemented with electron-electron interactions, then it is convenient to employ bosonization techniques \cite{Giamarchi2003}.
 Within a fixed $\hat{N}_F$ sector, we write $\hat{\psi}_\sigma(x) = (2\pi \xi)^{-1/2} e^{\iu \sigma(k_F - \pi/L)x} e^{\iu \hat{\theta}(x) - \iu \sigma \hat{\phi}(x)}$, where $k_F$ is the Fermi wavelength, $L$ is the length of the edge, $\xi$ is a short distance cutoff of the order $u/E_{\rm g}$ \cite{Maciejko2009} ($E_{\rm g}$ is the bulk gap), and $\hat{\phi}(x)$, $\hat{\theta}(x)$ are bosonic fields satisfying the commutation relations $[\hat{\phi}(x), \nabla \hat{\theta}(x')] = \iu \pi \delta(x-x')$. The edge modes are then described by the helical Luttinger liquid (HLL) theory \cite{Wu2006}
\begin{align}
\hat{H}_{\rm HLL} = \frac{u}{2\pi} \int \dif x\, \frac{1}{K} (\nabla \hat{\phi})^2 + K (\nabla \hat{\theta})^2
\label{eqHLLHam}
\end{align}
where $u$ is a velocity and the dimensionless Luttinger parameter $K$ quantifies the strength of interactions ($K < 1$ for repulsive interactions). In this description, TRS acts as $\hat{T}\hat{\phi}(x) \hat{T}^{-1} = \hat{\phi}(x) + \pi/2$; $\hat{T}\hat{\theta}(x) \hat{T}^{-1} = -\hat{\theta}(x) + \pi/2$ \cite{Hseih2020}.

More generally, the Hamiltonian \eqref{eqHLLHam} emerges as the long-wavelength limit of a renormalization group (RG) flow starting from a bare Hamiltonian that can feature many other TRS-respecting perturbations, e.g.~spin orbit coupling and spatial disorder. (We assume $K > 1/4$ throughout, which ensures that the HLL is stable with respect to the formation of a symmetry-broken insulator \cite{Maciejko2009}.) The usual bosonization identity relating bare electron operators and bosonic variables cannot be directly used in this case, since it neglects all spin texture in momentum space. Nevertheless, the renormalized fields $\hat{\phi}(x)$, $\hat{\theta}(x)$ obey the same symmetry properties as before.

It is well-known that the perfect conductance of these helical edge modes can be compromised if the electrons can tunnel into magnetic impurities \cite{Wu2006,Maciejko2009,Tanaka2011}, which induces an exchange coupling $J \sum_{\alpha=x,y,z}  \hat{S}^\alpha_{\rm el}(x) \otimes \hat{S}_{\rm imp}^\alpha$. The electron spin operators $\hat{S}^\alpha_{\rm el}(x) = \sum_{\sigma \sigma'} \hat{\psi}_\sigma^\dagger [\tau^{\alpha}]_{\sigma \sigma'}^{\vphantom{\dagger}} \hat{\psi}_{\sigma'}^{\vphantom{\dagger}}$ ($\tau^\alpha$ are the Pauli matrices) are odd under time-reversal \footnote{Strictly speaking, the exchange coupling is only an effective low-energy description of the physical Hamiltonian in which electrons tunnel between the system and the impurity. The operators acting on the system can therefore be thought of as breaking either TRS or fermion number conservation, both of which are required to protect the quantum spin Hall effect.}, and so can induce elastic backscattering even though TRS is preserved overall. We will instead consider electrostatic interactions between the HLL and a non-magnetic impurity, such that the Hamiltonian only features TRS-even, fermion-number-conserving operators acting on the system. Despite the absence of any tunnelling or exchange processes, we will show that this non-magnetic impurity can still give rise to quasi-elastic backscattering, leading to a deviation from quantized conductance that is in principle just as strong as a magnetic impurity.

For simplicity and concreteness, we model the impurity as a two-level system (TLS). We discuss possible physical manifestations of such TLSs below. For now, consider the impurity to have two low-energy configurations whose zero point energies differ by $E_{\rm sp}$, with a tunnelling matrix element $\Delta$ between the two \cite{Anderson1972}. The Hamiltonian is $\hat{H}_{\rm TLS} = (E_{\rm sp}/2) \hat{\sigma}^z + (\Delta/2) \hat{\sigma}^x$, where $\hat{\sigma}^{x,y,z}$ are the Pauli matrices in the impurity Hilbert space. In this basis, time reversal acts as $\hat{T} = K$, the complex conjugation operator.

Most generally, electrostatic interactions between the TLS and the quantum spin Hall system will take the form
\begin{align}
\hat{H}_{\rm int} = \int \dif^2 \vec{r}\, \hat{\rho}_{\rm el}(\vec{r}) \otimes \left[ \hat{\sigma}^xV_x(\vec{r}) + \hat{\sigma}^zV_z(\vec{r}) \right],
\label{eqCoupleBare}
\end{align}
where $\hat{\rho}_{\rm el}(\vec{r})$ is the density operator for the bare electrons, and $V_{x,z}(\vec{r})$ are arbitrary real functions of the 2D spatial coordinate $\vec{r}$, which we presume to be smooth on the scale of $\xi$, and localized near some point $x = 0$ along the edge. This interaction captures the dependence of both the splitting $E_{\rm sp}$ and tunnelling matrix element $\Delta$ on the distribution of electrons in the system. Note that the operators $\hat{\rho}_{\rm el}(\vec{r})$, $\hat{\sigma}^x$, and $\hat{\sigma}^z$ appearing in \eqref{eqCoupleBare} are all manifestly time-reversal invariant and fermion-number-conserving \footnote{More precisely, the system-impurity coupling operators we consider in this work are of the form $\hat{H}_{\rm int} = \sum_\alpha \hat{A}_\alpha \otimes \hat{B}_\alpha$, where $\hat{A}_\alpha$ and $\hat{B}_\alpha$ are Hermitian, TRS-respecting, and particle number conserving. The Hermiticity condition is required, otherwise operators of the form $\hat{A}_\alpha = \iu \hat{C}$, where $\hat{C}$ is TRS-odd (e.g.~a magnetic field), would be allowed, which obviously can induce backscattering \cite{Deng2020}.}.

\emph{Effective low energy theory.---} We now analyse the low-energy properties of the full Hamiltonian $\hat{H}_{\rm tot} = \hat{H}_{\rm HLL} + \hat{H}_{\rm TLS} + \hat{H}_{\rm int}$. In principle, an expression for the interaction \eqref{eqCoupleBare} in terms of the bosonic fields $\hat{\phi}(x)$, $\hat{\theta}(x)$ can be obtained, but for now we will simply ask which terms will generically arise by considering the symmetry properties of the operators acting on the system $\hat{A}_\alpha = \int \dif^2 \vec{r}\, \hat{\rho}_{\rm el}(\vec{r}) V_\alpha(\vec{r})$ ($\alpha = x, z$).
Evidently, $\hat{A}_\alpha$ are Hermitian, charge-conserving, TRS-invariant operators. Furthermore, since the interaction between the system and the TLS is localized around the point $x = 0$, we can perform a gradient expansion of the bosonic fields about this point (which is well-controlled at low energies), leaving only the fields $\hat{\phi}(x)$, $\hat{\theta}(x)$ and their spatial derivatives evaluated at $x = 0$. There are still infinitely many terms that meet these criteria, but for illustrative purposes we will consider just two
\begin{align}
\hat{H}_{\rm int} = J_z \nabla^2 \hat{\phi} \otimes \hat{\sigma}^z \;  + J_x:\!\nabla \hat{\theta} \cos[2\hat{\phi}]: \otimes\; \hat{\sigma}^x  + \cdots
\label{eqHamTerms}
\end{align}
with all fields evaluated at $x = 0$. (The colons denote normal ordering with respect to the product of $\nabla \hat{\theta}$ and $\cos[2\hat{\phi}]$.) The coefficients $J_{x,z}$ will depend in some complicated way on the microscopic details of the QSHE system in question as well as the profiles $V_{x,z}(\vec{r})$ in \eqref{eqCoupleBare}, but neither are constrained by time-reversal symmetry. In the Supplemental Material, we show how such terms arise from a microscopic Hamiltonian when spin-orbit coupling is explicitly taken into account.

To provide some intuition for the two perturbations considered here, if we were to map them back to fermionic operators using the bosonization identity then the first term would describe forward-scattering of electrons. The second term corresponds to single-particle backscattering between non-degenerate states, which is allowed because counterpropagating states of different energies are not related by TRS.

At tree level, both operators in Eq.~\eqref{eqHamTerms} are RG-irrelevant, with scaling dimensions $\Delta_1 = 2$, $\Delta_2 = 1 + K$. Therefore, the deviation from quantized conductance at leading order in $J_{x,z}$ will decrease as the temperature of the system is lowered as $T^{2K}$.
The reason we consider them here is that when perturbative loop corrections are included, the combination of these two operators can generate a new relevant operator
\begin{align}
\hat{H}_{y} = (y u/ \xi) \cos[2\hat{\phi}] \otimes \hat{\sigma}^y ,
\label{eqRelOp}
\end{align}
where $y$ is a dimensionless coupling constant. Specifically, the RG equation for $y$ takes the form
\begin{align}
\frac{\dif y}{\dif \ell} = (1-K)y - J_x J_z / u^2 \xi + \cdots
\label{eqRG}
\end{align}
where $\ell$ parametrizes the cutoff scale as $\xi = \xi_0 e^{\ell}$. The symmetries of the electrostatic coupling \eqref{eqCoupleBare} forbid a non-zero bare value of $y$, since $\cos[2\hat{\phi}]$ is a TRS-odd operator. Indeed, $\cos[2\hat{\phi}]$ generates elastic single-particle backscattering of the same kind that would be expected from a magnetic impurity \cite{Maciejko2009}. Nevertheless, we see that under RG a non-zero value of $y$ is generated if $J_x J_z \neq 0$. In fact, there are infinitely many combinations of operators that can appear in the bare Hamiltonian which contribute to $y$, as represented by the ellipsis in Eq.~\eqref{eqRG}. Evidently, the operator $\hat{H}_y$ describes quasi-elastic backscattering of electrons (involving an energy transfer of $\epsilon$) accompanied by a transition of the TLS.

Our derivation of the RG equation \eqref{eqRG} (described in detail in the Supplemental Material \cite{SM}) is a variant of the Anderson-Yuval-Hamann approach to the Kondo model, where the partition function for $\hat{H}_{\rm tot}$ is mapped onto a classical statistical mechanics problem with a single spatial dimension representing imaginary time \cite{Anderson1970}. Because the underlying theory \eqref{eqHLLHam} is free, we can compute the operator product expansion of the two terms in Eq.~\eqref{eqHamTerms}, from which the one-loop beta function can be obtained using standard methods (see e.g.~Ref.~\cite{Fradkin2013}).

One can develop a more intuitive understanding of how the elastic backscattering term \eqref{eqRelOp} arises by successively applying the two terms in Eq.~\eqref{eqHamTerms} in a perturbative manner, as illustrated in the inset of Fig.~\ref{figElectro} (working in a fermionic representation). An electron undergoes a transition proceeding via an intermediate virtual state, which will be either left- or right- moving depending on whether the forward- or back-scattering term ($J_z$ and $J_x$, respectively) is applied first. If the operators acting on the impurity were not included in Eq.~\eqref{eqHamTerms}, then the contributions from the two choices of ordering would destructively interfere due to TRS. However, because $\hat{\sigma}^{x}$ and $\hat{\sigma}^z$ anticommute, an additional relative phase of $\pi$ between the two contributions is introduced. The result is elastic backscattering accompanied by a transition of the state of the impurity.

Since $J_x$, $J_z$ are RG-irrelevant and $y$ is RG-relevant (or marginal for noninteracting fermions $K = 1$), for large $\ell$ the right hand side of \eqref{eqRG} will be dominated by the first term, giving $y(\ell) \approx y_0 e^{(1-K)\ell}$ where $y_0 \sim J_x(0) J_z(0) / u^2 \xi_0 $ is determined by the early stages of the RG flow. The low-energy properties of the system will therefore be equivalent to a system with the Hamiltonian $\hat{H}_{\rm eff} = \hat{H}_{\rm HLL} + \hat{H}_{\rm TLS} + (y_0 u / \xi)\, \cos[2\hat{\phi}] \otimes \hat{\sigma}^y$ (keeping only relevant operators). Because there are many other combinations of bare operators that contribute to Eq.~\eqref{eqRG}, and the parameters $J_x$, $J_z$ have not yet been determined in terms of microscopic parameters, we will treat $y_0$ as a phenomenological parameter.

\emph{Conductance.---} Having derived the effective low-energy theory, we can now calculate the resistance induced by the HLL-TLS interactions by adapting the standard derivation for the conductance of a Luttinger liquid in the presence of a static impurity \cite{Kane1992a}. At leading order in $y_0$, we find a linear DC resistance of
\begin{align}
\frac{R(T)}{h/e^2} &= \frac{\pi^2 y^2_0}{2}\left(\frac{2\pi T}{E_{\rm g}}\right)^{2K - 2} \hspace{-5pt}\sech\left(\frac{\epsilon}{2T}\right) \frac{|\Gamma(K + \iu \epsilon/2\pi T)|^2}{\Gamma(2K)}, \label{eqGDC}
\end{align}
where $h = 2\pi \hbar$, $\Gamma(x)$ is the Euler gamma function, and $\epsilon = \sqrt{E_{\rm sp}^2 + \Delta^2}$ is the difference in energy of the eigenvalues of $\hat{H}_{\rm TLS}$
(see the Supplemental Material for details \cite{SM}).
Eq.~\eqref{eqGDC} is valid when the RG flow \eqref{eqRG} is cut off before the coupling $\hat{H}_y$ becomes non-perturbatively strong, which occurs at an energy scale $E_y = E_{\rm g}y_0^{1/(1-K)}$. The RG may be cut off by nonzero temperature $T$, or by the splitting $\epsilon$. If $T \gtrsim \max(\epsilon, E_y)$, then the resistance scales as $R(T) \sim T^{2K-2}$, as could be anticipated from the scaling dimension of $y$. (In fact, in this limit Eq.~\eqref{eqGDC} tends to the same resistance as would be induced by a static Zeeman field.) If $\epsilon \gtrsim \max(T, E_y)$, then the impurity becomes frozen in its ground state. Backscattering is then suppressed, since the TLS cannot efficiently undergo a transition, and the contribution \eqref{eqGDC} to the edge resistance becomes thermally activated $R(T) \sim e^{-\epsilon/T}$. This reflects the quasi-elastic nature of the process described here. In this regime, the dominant source of resistance is due to inelastic scattering, giving $R(T) \sim T^\eta$, with $\eta = \min(2K+2, 8K-2)$ \cite{Lezmy2012}.

If instead $T, \epsilon \lesssim E_y$, then the RG flows to a strong coupling regime where the strength of the perturbation $yu/\xi$ becomes of the order of $E_{\rm g}$. Eq.~\eqref{eqGDC} then no longer applies. By analogy to the Kondo effect in helical liquids \cite{Maciejko2009}, we expect that the TLS will hybridize with bulk electrons, and the composite system (i.e.~the edge and bulk degrees combined with the TLS) can be thought of as an isolated QSH insulator with a gap $\sim E_y$, plays the r{\^o}le of the Kondo temperature in this problem. The remaining backscattering processes are then inelastic, and again we have $R(T) \sim T^\eta$.

\emph{Discussion.---} We have shown that electrostatic interactions between a dynamic impurity and helical electrons facilitate quasi-elastic backscattering, leading to an edge mode resistance that is potentially much stronger than previously studied inelastic backscattering mechanisms at low temperatures. The resistance increases (or remains constant for $K=1$) as the temperature is lowered, following a power law $T^{2K-2}$ down to a non-universal cutoff scale $E_{\rm cut} = \max(\epsilon, E_y)$. At temperatures below $E_{\rm cut}$, the TLS becomes frozen either by its own dynamics or by interactions with bulk electrons, and the resistance then scales as $T^\eta$ (see Fig.~\ref{figRes}).

The topological protection of helical edge modes is usually attributed to the fact that TRS-invariant Hamiltonian perturbations $\hat{H}$ cannot couple counterpropagating Kramers-degenerate states $\ket{\psi}$ and $\ket{\bar{\psi}}$, since $\braket{\bar{\psi}|\hat{H}|\psi} = 0$. However, such an argument only applies to situations where the quantum spin Hall insulator is isolated. Here, the system is coupled to additional degrees of freedom that make up the impurity. TRS then applies at the level of the composite system-plus-impurity, rather than just the system, and so transitions from $\ket{\psi}$ to $\ket{\bar{\psi}}$ are not forbidden, provided the environment undergoes a simultaneous transition. (See Ref.~\cite{McGinley2020} for a similar example in 1D symmetry-protected topological phases.)

\begin{figure}
	\includegraphics[scale=1]{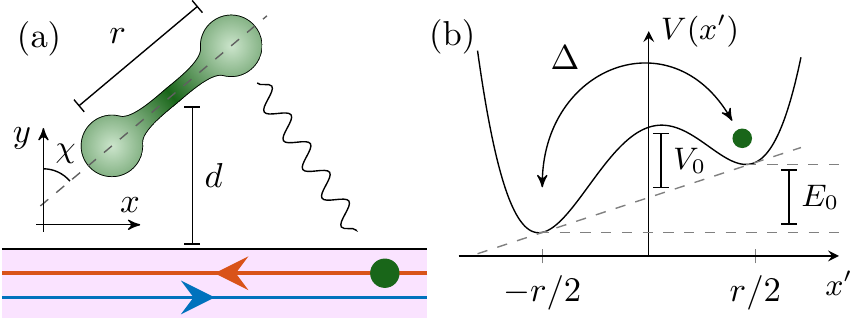}
	\caption{Possible realization of an impurity capable of inducing quasi-elastic backscattering. (a) An electron is confined within one of two potential wells (light green regions) a distance $d$ from the helical edge modes, oriented at an angle $\chi$ from the normal. $d$ may be much larger than a tunnelling length. (b) Potential energy profile along the double well axis $x'$. The tunnelling matrix element between the wells is $\Delta$.}
	\label{figWell}
\end{figure}

This same reasoning can be used to understand how quasi-elastic backscattering can occur in scenarios where helical electrons are tunnel-coupled to magnetic impurities \cite{Maciejko2009} or charge puddles tuned to resonance \cite{Vayrynen2013}. However, the key aspect that distinguishes the mechanism described here from those previous results is that it can occur even when tunnelling of electrons out of the edge modes is suppressed. As such, our analysis encompasses a much broader class of impurities -- they do not need to be within a tunnelling length $\ell_{\rm tun}$ of the sample, and they can be composed of either electrons or ions. For instance, the impurities could have the same origin as the effective two-level systems used to account for the ubiquity of `$1/f$ noise' in solid state systems \cite{Dutta1979,Weissman1988}. 

As a simple example of a backscattering-inducing impurity, consider an electron trapped in a double quantum well a distance $d$ from the helical edge. At sufficiently low temperatures, the electron will be in the ground state of one of the two wells, and hence can be treated as a TLS with $E_{\rm sp}$ and $\Delta$ determined by the potential energy landscape near the wells \cite{Anderson1972}; see Fig.~\ref{figWell}. Straightforward calculations (detailed in the Supplemental Material \cite{SM}) allow one to calculate the profiles $V_{x, z}(\vec{r})$, and in turn the effective strength of the backscattering $y_0^2$, which we find to scale as $d^{-4}$. This differs from backscattering mechanisms involving tunnelling out of the helical edge in that the dependence on $d$ is only algebraic, rather than exponential $\sim e^{-d/\ell_{\rm tun}}$. Similar power-law interactions should arise for more general types of impurity. Given that the backscattering mechanism here is significantly longer ranged than charge puddle-mediated effects, we expect our findings to be of particular relevance to systems with a band gap large enough to render the formation of nearby charge puddles unlikely, e.g.~in atomically thin crystals of WTe$_2$ and related compounds \cite{Wu2018}.

While we have adopted a two-level system model to describe the impurity, our arguments naturally generalise to multi-level impurities. One expects that the low-energy effective theory will generically contain all operators of the form \eqref{eqRelOp} in which $\hat{\sigma}^y$ is replaced with other Hermitian, TRS-odd operators. The crossover scale $\epsilon$  below which the impurity dynamics is frozen is again set by the energy difference between the two lowest eigenstates. For certain kinds of impurity, $\epsilon$ may be small or zero by symmetry, in which case the resistance persists down to correspondingly low temperatures. For instance, the low-energy states may describe two ionic configurations related by symmetry, giving $E_{\rm sp} = 0$, and hence $\epsilon = |\Delta|$. Alternatively, the impurity may be formed of an odd number of electrons, which gives $\epsilon = 0$ by Kramers theorem. Therefore, in regimes where $\epsilon$ is sufficiently small, our results are consistent with the weak temperature dependence of the edge mode resistance seen in experiment \cite{Fei2017}.

Finally, for the same general reasons, analogous effects should arise in other systems featuring topological modes where TRS plays an important r{\^o}le, e.g.~3D topological insulators and Dirac semimetals.

\begin{acknowledgements}
	\emph{Acknowledgements.---}This work was supported by an EPSRC studentship and Grants No.~EP/P034616/1 and No.~EP/P009565/1, and by an Investigator Award of the Simons Foundation.
\end{acknowledgements}

\bibliography{\jobname}

\begin{thebibliography}{36}
\bibitem [{Zha()}]{ZhaiNote}%
  \BibitemOpen
  \href@noop {} {}\bibinfo {note} {Backscattering can also be induced by a
  fluctuating field that couples to a time-reversal symmetry breaking
  observable \cite{Deng2020}.}\BibitemShut {Stop}%
\end{thebibliography}

\newpage

\clearpage


\appendix

\setcounter{figure}{0}
\makeatletter 
\renewcommand{\thefigure}{S\arabic{figure}}

\newcounter{defcounter}
\setcounter{defcounter}{0}

\newenvironment{myequation}
{%
	\addtocounter{equation}{-1}
	\refstepcounter{defcounter}
	\renewcommand\theequation{S\thedefcounter}
	\align
}
{%
	\endalign
}

\begin{onecolumngrid}
	\begin{center}
		{\fontsize{12}{12}\selectfont
			\textbf{Supplemental Material for ``\papertitle''\\[5mm]}}
		{\normalsize Max McGinley and Nigel R. Cooper\\[1mm]}
		{\fontsize{9}{9}\selectfont  
			\textit{\tcm}}
	\end{center}
	\normalsize
\end{onecolumngrid}

\vspace{20pt}

\begin{twocolumngrid}
	
\subsection*{RG Analysis}

Here we derive the renormalization group flow equation \eqref{eqRG} quoted in the main text. Our starting point is the Hamiltonian $\hat{H}_{\rm tot} = \hat{H}_{\rm HLL} + \hat{H}_{\rm TLS} + \hat{H}_{\rm int}$, where $\hat{H}_{\rm int}$ is given by \eqref{eqHamTerms}. We consider the partition function $\mathcal{Z} = \Tr e^{-\hat{H}_{\rm tot}/T}$ in the imaginary time ($\tau$) path integral representation. It is convenient to integrate out the $\theta(\tau,x)$ field using the operator equation of motion $\partial_\tau \hat{\phi} = \iu u K \nabla \hat{\theta}$, giving $\mathcal{Z} = \int \mathcal{D}\phi(\tau,x) \mathcal{D} \vec{S}(x,\tau) e^{-S[\phi(\tau,x), \vec{S}(\tau)]}$, where $\vec{S}(\tau)$ is the 3-component pseudospin describing the two level system, and the action is \cite{Giamarchi2003}
\begin{myequation}
S[\phi(\tau,x), \vec{S}(\tau)] &= S_0[\phi(\tau,x), \vec{S}(\tau)] + S_1[\phi(\tau,x), \vec{S}(\tau)] \nonumber\\
S_0[\phi(\tau,x), \vec{S}(\tau)] &\coloneqq \frac{1}{2\pi K}\int_0^\beta \dif \tau \int \dif x\; u(\partial_x \phi)^2 + \frac{1}{u}(\partial_\tau \phi) \nonumber\\ &+ S_{\rm WZ}\big[\,\vec{S}(\tau)\big] 
\end{myequation}
where $\beta = 1/T$, $S_1$ represents the terms coming from $\hat{H}_{\rm TLS} + \hat{H}_{\rm int}$, and $S_{\rm WZ}$ is the Wess-Zumino term \cite{Fradkin2013}. (We do not require an expression for the latter, since we will evaluate spin correlators in the operator formalism.) The term $S_1$ will be treated as a perturbation about the fixed point action $S_0$. Without loss of generality, we can write
\begin{myequation}
S_1\big[\phi, \vec{S}\;\big] &= \sum_{\nu} \sum_{\mu=0,x,y,z} g_{\nu \mu} \int_0^\beta \frac{\dif \tau}{\sigma^{1-\Delta_\nu}} A_\nu[\phi](\tau)\, S^\mu(\tau),
\label{eqActionPert}
\end{myequation}
where $S^0(\tau) \coloneqq 1/2$, and $A_\nu[\phi](\tau)$ are scaling operators depending on the field $\phi(x,\tau)$ at the time $\tau$, with scaling dimensions $\Delta_{\nu}$ \cite{Fradkin2013}. Here, $\sigma = \xi/u$ is a short-time cutoff of the order of the inverse bulk gap.

The partition function of the perturbed theory can be formally expanded in terms of expectation values with respect to $S_0$
\begin{widetext}
\begin{myequation}
\frac{\mathcal{Z}}{\mathcal{Z}_0} &= \sum_{n=0}^\infty (-1)^n \sum_{\substack{\nu_1 \ldots \nu_n\\ \mu_1 \ldots \mu_n}}\left(\prod_{a=1}^n g_{\nu_a \mu_a}\right)
\int_0^\beta \frac{\dif \tau_1}{\sigma^{1-\Delta_{\nu_1}}}
\cdots
\int_0^{\tau_{n-1} - \sigma} \frac{\dif \tau_n}{\sigma^{1-\Delta_{\nu_n}}}\,
\Braket{A_{\nu_1}(\tau_1) \cdots A_{\nu_n}(\tau_n)} \Braket{S^{\mu_1}(\tau_1) \cdots S^{\mu_n}(\tau_n)}
\label{eqPathIntegFormal}
\end{myequation}
\end{widetext}
where $\mathcal{Z}_0$ is the partition function for the action $S_0$. Here, we impose a short-time cutoff by demanding that the time coordinates $\tau_i$ are always separated by a time of at least $\sigma \sim E_{\rm g}^{-1}$. Although crude, this cutoff procedure is accurate enough to determine the one-loop beta function \cite{Fradkin2013}. Note that because the perturbation $S_1$ only acts at $x = 0$, the effective dimension for the problem is 0+1D.

This type of expansion forms the basis of a study of the Kondo model by Anderson, Yuval, and Hamann \cite{Anderson1970}. To help provide some physical intuition, they identified Eq.~\eqref{eqPathIntegFormal} with the grand partition function for a one-dimensional classical gas of particles interacting via long-range forces governed by the correlators of $A_\nu$ and $S^\mu$, subject to a hardcore constraint $|\tau_i - \tau_j| \geq \sigma$. In this analogous classical system, imaginary time $\tau$ plays the r{\^o}le of the spatial coordinate, the length of the system is $\beta$, and $g_{\nu \mu}$ are fugacities for the various `flavours' of particle, which we assume to be small. The Anderson-Yuval-Hamann RG scheme involves integrating out configurations in which two particles are separated by a distance $\sigma \leq \Delta t < b \sigma$, and then rescaling the coordinate $\tau$ by a factor $b^{-1}$ to restore the original cutoff $\sigma$.

For an infinitesimal RG step $b = 1 + \delta \ell$ in the dilute gas regime $g_{\nu \mu} \ll 1$, configurations in which more than two particles are separated by $|\tau_i - \tau_j| < b\sigma$ are rare enough to be neglected. The integration step can then be performed by replacing the two nearby particles by a single particle (possibly of a different flavour), chosen such that the potential felt by the other particles far away is unchanged. This effectively changes the fugacity of the new particle. In this limit, the appropriate renormalization of the $g_{\nu \mu}$ can be determined using the operator product expansion (OPE) formalism; see Refs.~\cite{Fradkin2013} for an introduction. The OPE of two scaling operators $A_{\nu}(\tau)$ $A_{\nu'}(\tau')$ acting on the system describes how their product behaves as the coordinates $\tau$, $\tau'$ approach one another, and takes the form
\newcommand{\sepvar}{\varsigma}
\begin{myequation}
&\lim_{\sepvar \rightarrow 0}\; \mathcal{T}_\tau A_{\nu}(\bar{\tau}+\sepvar/2) A_{\nu'}(\bar{\tau}-\sepvar/2) \nonumber\\=& \lim_{\sepvar \rightarrow 0}\; \sum_{\nu''} \frac{c_{\nu \nu'; \nu''}(\sepvar)}{ |\sepvar|^{\Delta_\nu + \Delta_{\nu'} - \Delta_{\nu''}}} A_{\nu''}(\bar{\tau})
\end{myequation}
where $\mathcal{T}_\tau$ denotes time ordering. Here the dimensionless functions $c_{\nu \nu'; \nu''}(\sepvar)$ are either constant in $\sepvar$ or proportional to $\sgn \sepvar$.

In an isolated system, the OPE for the system operators suffices to determine the one-loop RG equations. The symmetries of the fixed point action and the operators $A_\nu$, $A_{\nu'}$ are preserved under this process. (Note that we have been careful to avoid any spurious time-reversal symmetry breaking in the OPE by using the symmetrized time coordinate $\bar{\tau} = (\tau + \tau')/2$.). However, in our case the RG also depends on the OPE for the pseudospin fields $S^\mu(\tau)$, which we write as
\begin{myequation}
\mathcal{T}_\tau S^\mu(\bar{\tau}+\sepvar/2) S^{\mu'}(\bar{\tau}-\sepvar/2) = \sum_{\mu''} d_{\mu\mu'; \mu''}(\sepvar) S^{\mu''}(\bar{\tau}),
\end{myequation}
again to be understood in a weak sense. The coefficients $d_{\mu\mu'; \mu''}(\sepvar)$ can be evaluated in the operator representation
\begin{myequation}
d_{\mu\mu'; \mu''}(\sepvar) &= \frac{1}{4} \delta_{\mu\mu'} +  \frac{\iu}{2} \sgn(\sepvar) \epsilon_{\mu \mu' \mu''}
\label{eqSpinOPE}
\end{myequation}
for $\mu, \mu', \mu'' \in \{x,y,z\}$.
Generalising the arguments given in Ref.~\cite{Fradkin2013}, we find that after the infinitesimal RG step, $\mathcal{Z}$ will be left invariant if the fugacities $g_{\nu \mu}$ are renormalized by
\begin{myequation}
\frac{\dif g_{\nu \mu}}{\dif \ell} = (1-\Delta_\nu) g_{\nu \mu} - \sum_{\substack{\nu' \nu''\\ \mu' \mu''}} g_{\nu' \mu'} g_{\nu'' \mu''} f^{\mu' \mu''; \mu}_{\nu' \nu''; \nu}
\end{myequation}
where
\begin{myequation}
f^{\mu' \mu''; \mu}_{\nu' \nu''; \nu} &= \frac{1}{2}\big[c_{\nu' \nu''; \nu}(+\sigma) d_{\mu' \mu''; \mu}(+\sigma) \nonumber\\ &+ c_{\nu' \nu''; \nu}(-\sigma) d_{\mu' \mu''; \mu}(-\sigma) \big]
\label{eqRGCoeff}
\end{myequation}
These two terms correspond to the two different orders in which the operators $A_{\nu} S^\mu$ and $A_{\nu'} S^{\mu'}$ can appear. Because $d_{\mu' \mu''; \mu}(\sepvar)$ has a nontrivial dependence on $\sepvar$ for $\mu' \neq \mu''$, system operators that would be forbidden by TRS in an isolated system can actually generated under the RG. This is best illustrated using the two terms we introduced in Eq.~\eqref{eqHamTerms}. After substituting for $\hat{\theta}$, the scaling dimensions of the fields $\nabla^2 \phi$ and $\iu \partial_\tau \phi \cos[2\phi]$ are $\Delta_1 = 2$ and $\Delta_2 = 1 + K$, respectively. With proper normalization, the system operators in question can be written as $u^2 \sigma \nabla^2 \phi(\tau)$ and $\sigma^{-K} \iu \partial_\tau \phi \cos[2\phi]$, and the fugacities are $g_{1 z} = J_z/(u^2\sigma)$ and $g_{2 x} = -J_x/uK$. The OPE can be computed with the help of Wick's theorem
\begin{myequation}
&\big[u^2 \sigma \nabla^2\phi(\bar{\tau}+\sepvar/2)\big]  \bigg[ \sigma^{-K} :\iu \partial_{\tau} \phi(\bar{\tau}-\sepvar/2) \cos[2\phi(\bar{\tau}-\sepvar/2)]: \bigg] \nonumber\\
&= \frac{-\iu K}{\sepvar^3} \big[\sigma^{-K}\cos(2\phi(\bar{\tau}))\big] + \cdots
\end{myequation}
where we have omitted other less relevant operators. The above comes from the contribution in which the $\nabla^2 \phi$ operator is Wick contracted with $\iu \partial_\tau \phi$, which can be evaluated using the short-distance expression for the Green's function $\braket{\phi(x,\tau) \phi(0,0)} = (K/4)\log[x^2 + u^2 \tau^2] + \text{const.}$ \cite{Giamarchi2003}. Note that the coefficient is odd in $\sepvar$, and so without the TLS operators the two terms in \eqref{eqRGCoeff} would cancel. However, since $S^x$ and $S^z$ are non-commuting, the coefficients $d_{\mu' \mu''; \mu}(\sepvar)$ introduce an additional factor of $\sgn \sepvar$ [Eq.~\eqref{eqSpinOPE}], and so this cancellation does not occur. The elastic backscattering operator \eqref{eqRelOp} is therefore generated under the RG, even though the time-reversal symmetry of the electrostatic interactions forbids a non-zero bare value of the dimensionless coupling constant $y$. The scaling dimension for the above term is $\Delta = K$, and so RG equation for $y$ is
\begin{myequation}
\frac{\dif y}{\dif \ell} = (1-K)y - K g_{1 z} g_{2 x} + \cdots
\label{eqRGRel}
\end{myequation}
which upon substitution for $J_{x,z}$ gives Eq.~\eqref{eqRG}.

The perturbation $S_1$ includes the Hamiltonian $\hat{H}_{\rm TLS} = (E_{\rm sp}/2) \hat{\sigma}^z + (\Delta/2) \hat{\sigma}^x$. The dimensionless fugacities for these terms are $g_z = \sigma E_{\rm sp}/2$, $g_x = \sigma \Delta/2$, and both have scaling dimension $\Delta_{x,z} = 1$. Therefore, the dilute gas approximation $g_{\nu \mu} \ll 1$ will only be valid for $\ell$ up to the point where $g_{x,z}(\ell) \sim 1$, at which point these terms become non-perturbatively strong, and we must resort to methods that are exact in $E_{\rm sp}$, $\Delta$. As expected, the corresponding energy scale where this occurs is $E_{\rm g} e^{-\ell} \sim \epsilon$. The analogous scale where the system-impurity coupling \eqref{eqRelOp} becomes non-perturbatively strong is $E_y \coloneqq E_{\rm g} y_0^{1/(1-K)}$. Scaling behaviour governed by the fixed point $S_0$ can therefore be expected for $T \gtrsim E_{\rm cut} = \max(\epsilon, E_y)$.

\subsection*{Resistance at leading order in $y$}

Here we derive expressions for the electrical resistance of the helical Luttinger liquid using the effective low-energy Hamiltonian $\hat{H}_{\rm eff} = \hat{H}_{\rm HLL} + \hat{H}_{\rm TLS} + (y_0 u / \xi)\, \cos[2\hat{\phi}] \otimes \hat{\sigma}^y$ and working to lowest order in $y_0$. Our calculation is an extension of Kane and Fisher's derivation for the conductance of a Luttinger liquid coupled to a static impurity \cite{Kane1992a}. Again working in an imaginary time path integral formalism, we first integrate out $\hat{\theta}(x)$ over all $x$ and $\hat{\phi}(x)$ for all $x \neq 0$, leaving only $\hat{\phi}(x=0)$. After rotating the spin quantization axis to one in which $\hat{H}_{\rm TLS}$ is diagonal, one obtains the Matsubara action
\begin{myequation}
S_{\rm eff}\big[\phi,\, \vec{S}\;\big] &= S_{\rm WZ}\big[\,\vec{S}\;\big] + \frac{1}{\pi K} \sum_{\iu \omega_n} |\omega_n| |\phi(\iu \omega_n)|^2 \nonumber\\ &+ \int_0^\beta \dif \tau\, \epsilon S^z(\tau) + (y/\sigma) S^y(\tau) \cos[2\phi(\tau)]
\end{myequation}
where $\omega_n = 2\pi n/\beta$ are the bosonic Matsubara frequencies, $\epsilon = \sqrt{E^2 + \Delta^2}$, and $\phi(\iu \omega_n) = \int_0^\beta \dif \tau\, e^{\iu \omega_n \tau} \phi(\tau)$ is the frequency space representation of the field $\phi(\tau)$.
The field $\phi(\tau)$ couples to a classical gauge field $a(\tau)$ via $S_a[\phi] = \int \dif \tau j(\tau) a(\tau)$, where $j(\tau) = -\iu e \partial_\tau \phi(\tau) / \pi$ is the current operator \cite{Giamarchi2003}. The continuation of the gauge field to real time is related to the voltage applied across the point $x = 0$ by $V(t) = \partial_t a(t)$ \cite{Kane1992a}. The term $S_a[\phi]$ can then be removed by a shift of variables $\phi(\tau) \rightarrow \phi(\tau) - eK a(\tau)/2$, giving
\begin{myequation}
&S_{\rm eff}\big[\phi,\, \vec{S}\;\big] = S_{\rm WZ}\big[\,\vec{S}\;\big] + \frac{1}{\pi K} \sum_{\iu \omega_n} |\omega_n| |\phi(\iu \omega_n)|^2 \nonumber\\ &+ \int_0^\beta \dif \tau\, \epsilon S^z(\tau) + (y/\sigma) S^y(\tau) \cos[2\phi(\tau) - eK a(\tau)] \nonumber\\ &+ \frac{Ke^2}{4\pi} \sum_{\iu \omega_n} |\omega_n| |a(\iu \omega_n)|^2 
\label{eqActionGauge}
\end{myequation}
where $S_{\rm WZ}$ is the Wess-Zumino term, as in the previous section. 
The partition function for the action \eqref{eqActionGauge} can be expanded in powers of $y$, and at leading order we have
\begin{widetext}
\begin{myequation}
\mathcal{Z}[a] &= \mathcal{Z}_0[a]\left(1 + \frac{y^2}{4\sigma^2} \int_0^\beta \dif \tau_1 \int_0^\beta \dif \tau_2\, F^{\rm m}(\tau_1 - \tau_2) C^{\rm m}(\tau_1 - \tau_2)   \cos[eK a(\tau_1) - eK a(\tau_2)]\vphantom{\frac{y^2}{2\sigma^2} \int_0^\beta}  \right)
\end{myequation}
where we have defined the imaginary time correlators
\begin{myequation}
F^{\rm m}(\tau_1 - \tau_2) &\coloneqq \Braket{\mathcal{T}_\tau e^{2\iu \phi(\tau_1)} e^{-2\iu \phi(\tau_2)}}_0, & C^{\rm m}(\tau_1 - \tau_2) &\coloneqq \Braket{\mathcal{T}_\tau S^y(\tau_1) S^y(\tau_2)}_0.
\end{myequation}
Here,  $\mathcal{Z}_0[a]$ and $\langle\, \cdot \, \rangle_0$ are the partition function and expectation values with respect to the action \eqref{eqActionGauge} at zero coupling $y = 0$. The current in imaginary time is given by the derivative of the generating functional $I(\tau) = \delta \log \mathcal{Z}[a] / \delta a(\tau)$, giving
\begin{myequation}
I(\tau) = I_0(\tau) - \frac{eK y^2}{2\sigma^2} \int_0^\beta  \dif \tau'\, F^{\rm m}(\tau - \tau') C^{\rm m}(\tau - \tau') \sin[eK a(\tau) - eK a(\tau')]
\end{myequation}
where $I_0(\tau) = \delta \mathcal{Z}_0 /\delta a(\tau)$ is the contribution for the unperturbed HLL, which is responsible for the conductance of the clean system $G_0 = K e^2/h$. We now perform an analytic continuation to the Keldysh contour in real time $t = -\iu \tau$, which runs from $t' = -\infty$ to $t' = t$ and then back to $t' = -\infty - \iu \beta$ \cite{Kane1992a}. This gives a correction $\delta I(t) = I(t) - I_0(t)$ of
\begin{myequation}
\delta I(t) = \frac{-\iu eK y^2}{2\sigma^2} \int_{-\infty}^t  \dif t' \bigg[F^{>}(t - t') C^{>}(t-t') - F^{<}(t - t') C^{<}(t-t')\bigg] \sin (eK [a(t) - a(t')])
\label{eqCurrCorr}
\end{myequation}
\end{widetext}
where $F^{>(<)}(t)$ is the greater (lesser) Green's function, which for $t > 0$ can be obtained by analytically continuing the imaginary time Green's function $F^{\rm m}(\tau)$ to $\tau \rightarrow \pm \iu t$, with the $+$ sign for $F^>(t)$ [similar for $C^{>,<}(t)$]. By standard techniques, one finds \cite{Giamarchi2003}
\begin{myequation}
F^{>(<)}(t) &= e^{\mp \iu \pi K \sgn(t)} \left(\frac{\pi \sigma/ \beta}{\sinh(\pi |t|/\beta)}\right)^{2K}, \nonumber\\
C^{>(<)}(t) &= \frac{1}{4}\left[\cos(\epsilon t) \mp \iu \tanh\left(\beta \epsilon/2\right) \sin(\epsilon t)\right]
\label{eqSpinCorrelator}
\end{myequation}
To obtain the linear conductance, we expand to first order in the voltage $V(t) = V_0 e^{-\iu \omega_0 t}$ which gives
\begin{myequation}
&\delta I_{\rm lin}(t) = V_0 e^{-\iu \omega_0 t} \frac{-\iu e^2 K^2 y^2}{4\omega_0 \sigma^2} \int_0^\infty \dif t' \left(1 - e^{\iu \omega_0 t' }\right)  \nonumber\\ &\times \bigg[\sin(\pi K)\cos(\epsilon t') + \cos(\pi K) \tanh\left(\beta \epsilon/2\right) \sin(\epsilon t')  \bigg] \nonumber\\ &\times \left(\frac{\pi \sigma/ \beta}{\sinh(\pi t'/\beta)}\right)^{2K}
\end{myequation}
The above can be evaluated using the standard integral $\int_0^\infty \dif x\, e^{\iu q x} \sinh^{-2K}(x) = 2^{2K - 1} P_K(-q)$, where we define $P_K(q) \coloneqq B(K + \iu q/2, 1- 2K)$, with $B(a,b) = \Gamma(a) \Gamma(b) / \Gamma(a + b)$ the Euler beta function. Using the identity $\sin(\pi K) \pm \iu \cos(\pi K) \tanh(\beta \epsilon/2) \equiv \sech(\beta \epsilon/2) \sin(\pi K \pm \iu \beta \epsilon/2)$ and using $\delta R \approx -\delta G / G_0^2$ with $G_0 = Ke^2/h$ \cite{Kane1992a}, we find the residual AC resistance within the linear response regime
\begin{widetext}
\begin{myequation}
&\delta R(\omega) = \frac{2\pi \hbar}{e^2} \times \frac{\iu\pi y^2}{4\omega \sigma} \left(\frac{2\pi \sigma}{\beta}\right)^{2K-1} \sech(\beta \epsilon/2)
\nonumber\\ &
\times
\bigg[\big(P_K[\epsilon \beta/\pi] - P_K[(\epsilon - \omega)\beta/\pi]\big)\sin(\pi K + \iu \beta \epsilon/2)
+ \big(P_K[-\epsilon \beta/\pi] - P_K[(-\epsilon-\omega)\beta/\pi]\big)\sin(\pi K - \iu \beta \epsilon/2) \bigg].
\label{eqResACGen}
\end{myequation}
Eq.~\eqref{eqResACGen} can be evaluated in the DC limit $\omega \rightarrow 0$, giving Eq.~\eqref{eqGDC}.
Additionally, we can compute the AC resistance at zero temperature. The real and imaginary parts are
\begin{myequation}
\Re \delta R_{T=0}(\omega) &= \begin{dcases}
0 & |\omega| < \epsilon, \\
\frac{2\pi \hbar}{e^2} \times \frac{\pi^2 y^2}{4\sigma |\omega|} \frac{(\sigma[|\omega| - \epsilon])^{2K-1}}{\Gamma(2K)} & |\omega| \geq \epsilon;
\end{dcases} \tag{\theequation a} \\
\Im \delta R_{T=0}(\omega) &= -\frac{2\pi \hbar}{e^2} \times \frac{\pi^2 y^2}{4\sigma \omega} \Gamma(1 - 2K)  \begin{dcases}
(\sigma[\epsilon+\omega])^{2K-1} + (\sigma[\epsilon-\omega])^{2K-1} - 2(\sigma \epsilon)^{2K-1} & |\omega| < \epsilon, \\
(\sigma[|\omega|+\epsilon])^{2K-1} - \cos(2\pi K) (\sigma[|\omega|-\epsilon])^{2K-1} - 2(\sigma\epsilon)^{2K-1} \hspace*{-5pt} & |\omega| \geq \epsilon.
\end{dcases}
\tag{\theequation b}
\label{eqGImAC}
\end{myequation}
\end{widetext}

We can also obtain an expression for the nonequilibrium current at finite DC bias by substituting $a(t) - a(t') = V\times (t-t')$ in Eq.~\eqref{eqCurrCorr}. At zero temperature, this gives
\begin{myequation}
&\delta R_{T=0}(V)\nonumber\\  &=  \frac{2\pi \hbar}{e^2} \times \begin{dcases}
0 & eV \leq \epsilon/K, \\
\frac{\pi^2 y^2}{4eK|V|\sigma} \frac{(\sigma[eK|V|-\epsilon])^{2K-1}}{\Gamma(2K)} & e|V| > \epsilon/K.
\end{dcases}
\label{eqGNonlin}
\end{myequation}

We note that in the limit $\epsilon \ll \max(T, \omega, eV)$, our expressions for the resistance exactly coincide with those of Kane and Fisher for a Luttinger liquid coupled to a static impurity, i.e.~a perturbation of the form $(y/\sigma) \cos[2\hat{\phi}]$, without the pseudospin operator. This is because the correlator $C^{>,<}(t)$ [Eq.~\eqref{eqSpinCorrelator}] becomes time-independent in this regime, and so the system behaves as if a TRS-breaking static magnetic impurity were present.

\subsection*{Microscopic model}

Here we derive relations between the phenomenological parameters $y_0$, $\epsilon$ referred to in the main text and microscopic quantities for a concrete physical system. The model we have in mind is that of an electron trapped within a double quantum well, interacting with a quantum spin Hall insulator via electrostatic interactions, as illustrated in Fig.~\ref{figWell}a. The axis joining the two wells is oriented at an angle $\chi$ from the normal of the quantum spin Hall insulator boundary. The potential landscape felt by the electron within the double well can be characterized by a distance $r$ separating the two minima; a bias energy $E_{0}$; energies of zero-point motion $\hbar \omega_{1,2}$ within the two minima; and a barrier height $V_0$ measured with respect to the average of the two well energies (see Fig.~\ref{figWell}b). The tunnelling matrix element $\Delta$ is of the order $\hbar \omega_0 e^{-\lambda}$, where $\omega_0 \sim \sqrt{V_0/m_er^2}$ is of the same order as $\omega_{1,2}$ ($m_e$ is the electron mass), and $e^{-\lambda}$ is the overlap between the ground state wavefunctions of each well, given by \cite{Anderson1972}
\begin{myequation}
	\lambda \approx \frac{1}{2}\left(\frac{2m_e V_0}{\hbar^2}\right)^{1/2} r.
\end{myequation}
At temperatures $k_B T \ll \hbar \omega_0$, the impurity can be accurately modelled as a two level system, corresponding to the ground states of each well. The Hamiltonian is $\hat{H}_{\rm TLS} = (E_{\rm sp}/2) \hat{\sigma}^z + (\Delta/2) \hat{\sigma}^x$, where $E_{\rm sp} = E_0 + \hbar(\omega_{2} - \omega_1)/2$. 

The potential landscape will be modified by the presence of an electron in the helical liquid due to Coulomb repulsion. Within the two-level system description, this leads to an effective alteration of $E_{\rm sp}$ and $\Delta$. For $r \ll d$, the change in the bias $E_0$ is approximately $-r \vec{\nabla}V_{\rm C} \cdot \vec{n}_\chi$, where $\vec{n}_\chi$ is a unit vector pointing from well 1 to 2, and $V_{\rm C}$ is the Coulomb potential. Similarly, the change in the barrier height $V_0$ is $-r^2 [\vec{n}_\chi \cdot \vec{\nabla}]^2 V_{\rm C}$. Ignoring the motion of the helical electron in the $y$ direction, we can obtain functions $V_x(x)$, $V_z(x)$, which determine the insulator-impurity interaction via Eq.~\eqref{eqCoupleBare}. Specifically, we have
\begin{myequation}\label{eqProfiles}
	V_z(x) &= r \frac{e^2}{4\pi\epsilon_0} \frac{d \cos \chi - x \sin \chi}{(d^2 + x^2)^{3/2}} \tag{\theequation a} \\
	V_x(x) &= \frac{r^3 }{4} \frac{e^2}{4\pi \epsilon_0} \frac{\Delta}{V_0} \left(\frac{2m_eV_0}{\hbar^2}\right)^{1/2} \nonumber\\ &\times
	\frac{3(d\cos \chi + x \sin \chi)^2 -d^2 - x^2}{(d^2 + x^2)^{5/2}} \tag{\theequation b}
\end{myequation}
We must now express the electron density operator $\hat{\rho}_{\rm el}$ in terms of the  bosonized fields. For simplicity, we assume that the helical edge is clean, and features homogeneous spin-orbit coupling. In this case, we can approximate \cite{Hseih2020}
\begin{myequation}
	\hat{\rho}_{\rm el}(x) \approx \nabla \hat{\phi}(x) + \zeta : \nabla \hat{\theta}(x) \cos[2\hat{\phi}(x)]: +\, \mathcal{O}(\zeta^2)
	\label{eqRhoPhi}
\end{myequation}
where the dimensionless constant $\zeta$ is of the order $k_F \ell_{\rm so}^2/\xi$ ($k_F$ is the Fermi wavevector and $\ell_{\rm so}$ is the spin orbit length, i.e.~the inverse of the momentum scale over which the spin quantization axis changes appreciably). We emphasise that if the helical liquid is subject to additional TRS-respecting perturbations, such as a disordered scalar potential, then a different relationship will apply.

After substituting Eqs.~(\hyperref[eqProfiles]{S23}, \ref{eqRhoPhi}) into \eqref{eqCoupleBare}, we employ a gradient expansion of the bosonic fields about the coordinate $x = 0$ at which the double well is closest to the helical edge. This will converge well at sufficiently low energies. For instance, a term of the form $\int \dif x\, V_z(x) \nabla \hat{\phi}(x)$ is approximated by $V_z^{(0)} \nabla \hat{\phi} + V_z^{(1)}\nabla^2 \hat{\phi} + \cdots$, where all fields are evaluated at $x = 0$, and $V_{x,z}^{(n)} \coloneqq \int \dif x\, x^n V_{x,z}(x)$. Here we will be more precise than in the main text, and keep track of the eight most relevant coupling terms, namely
\begin{myequation}
	\hat{H}_{\rm int} &= J_{f0z} \nabla \hat{\phi} \otimes \hat{\sigma}^z \;  + J_{b0z}:\!\nabla \hat{\theta} \cos[2\hat{\phi}]: \otimes\; \hat{\sigma}^z \nonumber\\
	&+ J_{f1z}\nabla^2 \hat{\phi} \otimes \hat{\sigma}^z \;  + J_{b1z}:\!\nabla^2 \hat{\theta} \cos[2\hat{\phi}]: \otimes\; \hat{\sigma}^z \nonumber\\
	&+(z \leftrightarrow x)
	\label{eqHamTerms2}
\end{myequation}
A generalization of the methods used to derive Eq.~\eqref{eqRG} gives us an RG equation for $y$ of the form
\begin{myequation}
	\frac{\dif y}{\dif \ell} &= (1-K)y - \frac{1}{u^2 \xi}\big[J_{f1z} J_{b0x} - J_{f0z} J_{b1x} \nonumber\\ &- J_{f1x} J_{b0z} + J_{f0x} J_{b1z}\big] 
	\label{eqRG2}
\end{myequation}
Note that each of the coefficients in \eqref{eqHamTerms2} are either even or odd under inversion symmetry, and the elastic backscattering operator \eqref{eqRelOp} is inversion symmetry even. This restricts which pairs of terms can contribute to \eqref{eqRG2}. For this reason, it is necessary to consider a scenario with a geometry that breaks inversion symmetry is broken (i.e.~$\chi \neq p\pi/2$ for $p \in \mathbb{Z}$), or to include inversion-symmetry-breaking terms in the decoupled Hamiltonian. We do not consider the latter scenario here, and so our expressions for the resistance will vanish at appropriate values of $\chi$, but we expect that in a more generic system, backscattering will still occur even for $\chi = p\pi/2$.

Again for $\ell \gg 1$, the solution of \eqref{eqRG2} is given by $y(\ell) = y_0 e^{(1-K)\ell}$, and here $y_0$ is proportional to $\zeta(V_z^{(0)}V_x^{(1)} - V_x^{(0)}V_z^{(1)}) / u^2 \xi$. Note that this value vanishes if $V_z(x) \propto V_x(x)$, since in this limite the system-impurity coupling \eqref{eqCoupleBare} can be written in a factorized form \cite{McGinley2020}. Using \eqref{eqGDC}, the resistance becomes of the order
\begin{myequation}
\frac{R(T)}{h/e^2} &\propto \left(\frac{e^2}{4\pi \epsilon_0 d}\right)^4 r^8\cos^4 \chi \sin^2 \chi \left(\frac{2\pi k_B T}{E_{\rm g}}\right)^{2K-2} \nonumber\\ &\times \frac{m_e\Delta^2}{\hbar^2 V_0} \frac{k_F^2 \ell_{\rm so}^4}{u^4\xi^4} \times f(\epsilon/k_BT)
\label{eqResMicro}
\end{myequation}
where $\epsilon = \sqrt{E_{\rm sp}^2 + \Delta^2}$, and $f(x) = \sech(x/2)|\Gamma(K + \iu x / 2\pi)|^2/\Gamma(K)^2$ is close to 1 for $x \ll 1$, and decays as $e^{-x}$ for $x \gg 1$. As highlighted in the main text, the dependence on $d$ is only power-law. (Note that $V_{z}^{(1)}$ has a logarithmic IR divergence which we have ignored here; it is likely to be cut off by the Thomas-Fermi length $\ell_{\rm TF}$, giving an extra overall factor of $[\log(\ell_{\rm TF}/d)]^2$.)

We note that the gradient expansion used to derive \eqref{eqResMicro} may not be valid at higher temperatures, in which case a different 1+1 dimensional RG scheme should be employed that can account for the full spatial profile of $V_{x,z}(x)$. We leave this to future work.

\end{twocolumngrid}

\end{document}